\documentclass[12pt]{article}
\usepackage{amssymb,euscript,epsf,graphicx}
\usepackage[hang,small,bf]{caption}

\newcommand{\be}[3]{\begin{equation}  \label{#1#2#3}}
\newcommand{\ee}{\end{equation}}
\newcommand{\ba}{\begin{array}}
\newcommand{\ea}{\end{array}}

\let\Large=\large
\let\large=\normalsize

\newsavebox{\uuunit}
\sbox{\uuunit}
    {\setlength{\unitlength}{0.825em}
     \begin{picture}(0.6,0.7)
        \thinlines
        \put(0,0){\line(1,0){0.5}}
        \put(0.15,0){\line(0,1){0.7}}
        \put(0.35,0){\line(0,1){0.8}}
       \multiput(0.3,0.8)(-0.04,-0.02){10}{\rule{0.5pt}{0.5pt}}
     \end {picture}}
\newcommand {\unity}{\mathord{\!\usebox{\uuunit}}}
\def\X{{\hbox{\tiny $X$}}}
\def\Y{{\hbox{\tiny $Y$}}}
\def\W{{\hbox{\tiny $W$}}}
\def\Z{{\hbox{\tiny $Z$}}}

\begin{document}


\begin{flushright}
\hfill{HU-EP-01/60}\\
\hfill{hep-th/0112136}

\end{flushright}

\vspace{15pt}

\begin{center}{ \Large{\bf 
Vacua of $N=2$ gauged supergravity derived \\[3mm]
from non-homogeneous quaternionic spaces 
}}

\vspace{30pt}

{
{\bf Klaus Behrndt}\footnote{E-mail: behrndt@physik.hu-berlin.de} 
\quad and \quad
{\bf Gianguido Dall'Agata}\footnote{E-mail: dallagat@physik.hu-berlin.de}
 }
\vspace{15pt}

{\it Humboldt Universit\"at zu Berlin,
Institut f\"ur Physik,\\ 
Invalidenstrasse 110, 10115 Berlin,
 Germany}
\vspace{80pt}   
  
{ABSTRACT}
\end{center} 

We discuss a class of 4-dimensional non-homogeneous quaternionic
spaces, which become the two known homogeneous spaces (${\mathbb
E}AdS_4$ and $SU(2,1)/U(2)$) in certain limits.  These moduli spaces
have two regions where the metric is positive definite, separated by a
non-physical region where the metric has timelike directions and which
contains a curvature singularity.  They admit four isometries and we
consider their general Abelian gauging.  The critical points of the
resulting superpotential and hence the possible domain wall solutions
differ significantly in the two regions.  On one side one can
construct only singular walls, whereas in the other we found a smooth
domain wall interpolating between two infra-red critical points
located exactly on the boundary of the physical allowed parameter
region.

\newpage


\section{Introduction}


Gauged supergravity has attracted much attention due to its relevance
for dual descriptions of supersymmetric field theories.  Especially
for the case with maximal supersymmetry, this correspondence seems to
be on solid grounds. But also if the supersymmetry is partially broken
reliable results may still be possible and the ultimate hope would be
to describe $D=4, N=1$ super Yang Mills by $D=5, N=2$ gauged
 supergravity.

The main effort in this program has been concentrated along two
directions.  One was the description of the renormalization group (RG)
flow in supergravity, see \cite{410, 430,GPPZ}, and the other is known as
the brane world scenario \cite{340, 350, 220}. In both cases the
supergravity solution is given by a domain wall, which can be
seen as a source (singular space) or as a smooth soliton. The latter
case necessarily requires the existence of two (connected) extrema of
the potential, whereas in the first case the source cuts off part of
the spacetime and both sides can be identified (in ${\mathbb Z}_2$
symmetric way). The source point of view appears natural in flux
compactification of string or M-theory \cite{460, 330, 300, 310,
490, 420, 230}.

At a given critical point the spacetime can become anti deSitter
($AdS$) and by looking at the behaviour of the warp factor close to
these points one can distinguish between two types of extrema.  If the
warp factor is exponentially large one calls it an UV extremum and if
the warp factor appears to be exponentially small it is an IR
extremum, see \cite{320, 290, 250}. This notation is obviously related
to the RG-flow application and an example for a flow interpolating
between UV and IR extrema was given in \cite{430, flatdw}.  For brane
world scenarios it is essential to have an exponential suppression on
both sides of the wall and hence, in order to build up a brane world
as a smooth soliton, it is important to have a potential with two IR
critical points. But so far the appropriate potential within gauged
supergravity could not be constructed. In $N=1,D=4$ supergravity it is
straightforward to write down the corresponding potential \cite{390},
but it could not been obtained from a certain gauging and hence cannot
be applied to 5-dimensional supergravity.
In fact, there was a critical discussion about
problems in realizing brane worlds in supergravity \cite{320, 290,
400, 470, 210}. To circumvent these no-go theorems was one of
the motivations for this paper. So let us discuss the issue in more
detail.

In $D=5, N=2$ gauged supergravity scalar fields enter vector, tensor
and hyper multiplets. If the superpotential does not depend on
hyperscalars, there are no IR critical points \cite{320, 290,
flatdw,math}.  This means that if we gauge only a subgroup of the
R--symmetry \cite{480} or gauge isometries of the vector multiplet
moduli space \cite{360, 370} the potential can have at most UV
critical points. Therefore, in order to obtain IR critical points it
is necessary to gauge an isometry of the hypermultiplet moduli
space. Supersymmetry requires that it is a quaternionic space
\cite{090} and the general couplings for the 5--dimensional case have
been worked out in \cite{130}. But even after including
hypermultiplets, the number of critical points is highly restricted
\cite{280, 270, flatdw}. If one for instance considers the theory of
gravity coupled only to hypermultiplets no multiple IR critical points
have been found. Single IR critical point are derived in \cite{430,
270, flatdw} and could also be obtained from sphere compactification
in \cite{380}.  This seems to be in agreement with expectations
motivatived by Morse theory that as long as the considered scalar
manifolds are topological trivial, which is the case for the classical
homogeneous moduli spaces, the number of critical points is highly
restricted \cite{260}.  Moreover, in \cite{flatdw,math} it was shown
that, when the scalar manifold of the hypermultiplets is chosen to be
symmetric, the set of critical points of the scalar potential,
obtained by gauging any of its isometries, is connected.  When the
matter couplings are more general it was shown \cite{flatdw} that one
can obtain multiple critical points with at least one IR direction,
but still no solution connecting them was possible.

With this motivation in mind, we discuss in this paper a class of
non--homogeneous 4-dimensional quaternionic spaces and the vacua that
one can obtain by gauging its isometries, for a related discussion
using harmonic superspace see \cite{111, 333}. Since the quaternionic
space plays a central role in this paper, let us add some more
remarks. Quaternionic spaces are special Einstein spaces, that allow
for a triplet of covariantly constant complex structures. They can
have positive or negative curvature, but as moduli spaces of hyper
multiplets only the negatively curved spaces appear.  If every point
on the manifold can be reached by acting with the isometry group they
are called homogeneous and these quaternionic spaces have been
classified some times ago \cite{200,190}. In 4 dimensions e.g., there
are only two symmetric spaces given by the cosets ${SO(4,1) \over
SO(4)}$ and ${SU(2,1) \over SU(2) \times U(1)}$. The first one is the
Euclidean anti deSitter space (${\mathbb E}AdS_4$) and the latter case
is the non-compact version of the complex projective space.  There are
of course also quaternionic spaces that are not homogeneous and,
similar to the vector multiplet moduli space, the {\em quantum} moduli
space for hypermultiplets is expected to be non--homogeneous, see also
\cite{510, 500, 501}.

The paper is organized as follows. After giving a short resume of
quaternionic geometry (see also the appendix of \cite{450} for the 4d
conventions), we will construct in section 3 a class of
non--homogeneous quaternionic spaces, which appear to be a special case
of known Einstein spaces \cite{book}. In section 4 we discuss the
structure of its isometries and the gauging of a general abelian
combination.  In section 5 we investigate the critical points of the
superpotential and solve the flow equations for an explicit
example. We will end this paper with a summary of our results.


\section{N=2 gauged supergravity and quaternionic geometry}


For the applications proposed in this paper we need to recall the
general features of the five--dimensional supergravity theory coupled
to an arbitrary number of hypermultiplets with a special attention to
the quaternionic structure of the scalar manifold.

The bosonic sector of $5D$, ${\cal N} =2$ supergravity coupled to
$n_H$ hypermultiplets has as independent fields: the f\"{u}nfbein
$e_\mu ^a$, the graviphoton $A_\mu$ with field strength $F_{\mu\nu}$
and the $4 n_H$ `hyperscalars' $q^X$.  The complete form for the
action and transformation laws can be obtained from \cite{130}, where the
allowed couplings and gaugings for gravity with all the short matter
multiplets of $ {\cal N} =2$ supersymmetry in five dimensions have
been worked out.  We repeat here only the main ingredients.  The
bosonic part of the Lagrangean is
\begin{eqnarray}
    \label{Lagrangean}
e^{-1} {\cal L}^{{\cal N}  = 2}_{\rm bosonic} &=& - \frac{1}{2} R
-\frac{1}{4} F_{\mu \nu } F^{\mu \nu } -\frac{1}{2} g_{\X\Y} 
{\cal D}_\mu q^{\X} {\cal D}^\mu q^{\Y}   - g^2 \,
 {\cal V}(\phi,q)+ \nonumber\\
&&+\frac{1}{6\sqrt{6}}
e^{-1}\varepsilon^{\mu\nu\rho\sigma\tau} F_{\mu\nu}
F_{\rho\sigma} A_\tau \,,
\end{eqnarray}
where the scalar potential is given by
\begin{equation}
     \label{potential}
{\cal V}  =  - 4 P^r P^r  + \frac34 k^{\X} k^\Y g_{\X\Y}(q)\,.
\end{equation}
The covariant derivative on the hyperscalars is defined as
\begin{equation}
{\cal D}_\mu q^\X=\partial _\mu q^\X+g\,A_\mu k^\X(q)\,,
 \label{calDbos}
\end{equation}
where $k^\X(q)$ is the Killing vector of the gauged isometry on the
quaternionic scalar manifold parameterized by the hyperscalars $q^\X$
and $P^r(q)$ is the corresponding prepotential given below.

In $ {\cal N} =2$ supergravity theories in 4,5 or 6 dimensions, the
manifold parameterized by the hyperscalars has a quaternionic
K\"{a}hler structure.  This is determined by the $4n_H$--bein
$f^{iA}_\X$ (as one--forms $f^{iA}=f^{iA}_\X dq^\X$), with the $SU(2)$
index $i=1,2$ and the $Sp(2n_H)$ index $A=1,\ldots,2n_H$, raised and
lowered by the symplectic metrics $C_{AB}$ and $\varepsilon_{ij}$.  By
definition, these manifolds have a metric which is given by
\begin{equation}
g_{\X\Y}\equiv  
f_{\X}^{iA}f_{\Y}^{jB}\varepsilon_{ij}C_{AB}=f_{\X}^{iA}f_{\Y iA}\,.
\label{defgXY}
\end{equation}
The vielbeine are covariantly constant, including the torsionless
Levi--Civita connection $\Gamma_{\X\Z}{}^\Y$, the
$Sp(2n_H)$ connection $\omega _\X{}^B{}_A$ and the
$SU(2)$ connection $\omega _{\X i}{}^j$, which are all functions of the
hyperscalars:
\begin{equation}
\partial _\X f_\Y^{iA}-\Gamma _{\X\Y}{}^\Z f_\Z^{iA}
+f_\Y^{iB}\omega  _{\X B}{}^A
  +\omega _{\X k}{}^i f_\Y^{kA}=0\,. \label{covconstf}
\end{equation}
Of course, to be quaternionic, these manifolds admit a triplet of complex
structures  ($r=1,2,3$)
\begin{equation}
J^r_\X{}^\Y \equiv - {\rm i} \, f_\X^{iA} (\sigma^r)_i{}^j f^\Y_{jA}
\end{equation}
that obey the quaternionic algebra
\begin{equation}
J^r J^s = - \unity_{4n_H} \delta^{rs} + \epsilon^{rst} J^t.
\end{equation}
An important object for our applications is the $SU(2)$ curvature,
which is defined by
\begin{equation}
{\cal R}_{\X\Y i}{}^j=2\partial _{[\X}\omega_{\Y]i}
{}^{j}-2\omega_{[\X|i|}{}^k\omega_{\Y]k}{}^j={\rm i} {\cal 
R}^r_{\X\Y}(\sigma _r)_i{}^j\ ,\qquad
{\cal R}^r = d \omega^r - \varepsilon^{rst} \omega^s \omega^t
 \label{cRXYij}
\end{equation}
with real ${\cal R}^r_{\X\Y}$.

For $n_H > 1$ one can prove that these manifolds are Einstein and
that the $SU(2)$ curvatures are proportional to the complex structures:
\begin{eqnarray}
\label{Ricci}
R_{\X\Y} &=& \frac{1}{4 n_H} g_{\X\Y} R \ ,\\
\label{scalar}
R &=& 4 n_H (n_H + 2) \nu \ , \\
\label{SU2}
{\cal R}^r_{\X\Y} &=& \frac12 \nu J^r_{\X\Y}
\end{eqnarray}
where $\nu$ is a coefficient which is fixed by supergravity to be
$\nu = -1$.

In the case  that $n_H = 1$ the above equations become part of
the definition of a quaternionic K\"ahler manifold.
The same constraints can also be expressed as the requirement that
the Weyl tensor of the manifold is (anti)self--dual \cite{090}
\begin{equation}
{}^\star{\cal  W} = \pm {\cal W}
\end{equation}

The Killing vectors of the quaternionic space $k_I^\X$  (the index $I$
labels now the different isometries) are related to an
$SU(2)$ triplet of real prepotentials $P^r(q)$ that are defined by the
relation~\cite{Galicki:1987ja,D'Auria:1991fj,100,130,450}:
\begin{equation}
{\cal R}_{\X\Y}^r k_I^\Y= D_\X P_I^r \,,\qquad D_\X P_I^ r\equiv
\partial_\X P_I^r+2\varepsilon^{rst}\omega _\X^sP_I^t \,.\label{defprep}
\end{equation}
These can be uniquely solved both for the Killing vectors
\begin{equation}
k^\Z_I=-\frac43 {\cal R}^{r\,\Z\X}  D_\X P_I^r 
 \label{KXandDP}
\end{equation}
or the prepotentials:
\begin{equation}
  P^r_I=  \frac{1}{2n_H}D_\X k_{I\Y}{\cal R}^{\X\Y r}\,.
 \label{valuePrI}
\end{equation}

As we will be interested not only in constructing non-homogeneous
quaternionic K\"ahler manifolds, but also in studying their vacua and
possible supersymmetric flows, we will also need the bosonic part of
the supersymmetry transformations of the fermions.  For vanishing
vectors, these are given by
\begin{eqnarray}
\delta _\epsilon \psi _{\mu i}  & = & { D}_\mu (\omega )\epsilon
_i+{\rm i} \frac{1}{\sqrt{6}}g\gamma _\mu P_{ij}\epsilon^j=
\partial_\mu  \epsilon _i+ \frac14 \gamma^{ab} \omega_{\mu \,
ab}\epsilon_i
- \omega_{\mu i} {}^{j}\epsilon_j +{\rm i}
\frac{1}{\sqrt{6}}g\gamma _\mu P_{ij}\epsilon ^j\,, \nonumber\\
\delta_\epsilon \zeta^A &=& -{\rm i}\frac{1}{2} f_{\X i}^A(\not\!{\partial}
q^{\X})\epsilon ^i + \frac{\sqrt{6}}{4} \,g\, \epsilon^i f_{\X i}^A 
\,k^\X\,.
 \label{transfos1}
\end{eqnarray}

Using the above--mentioned properties of the Killing vectors,
we can now introduce the scalar `superpotential' function $W$, that can be
read off the gravitino supersymmetry transformation, by
\begin{equation}
W=\sqrt{\frac13 P_{ij}P^{ij}}=\sqrt{\frac23 P^rP^r}\,,
\label{Wmydef}
\end{equation}
such that the potential gets the form:
\begin{equation}
 {\cal V}= - 6 W^2+\displaystyle \frac92 g^{\X\Y}\partial_\X
 W\partial_\Y W\, \label{pot}.
\end{equation}

For domain wall solutions that respect four--dimensional Poincar\'e
invariance we can write the metric as
\begin{equation}
ds^2 = e^{2 U(\rho)} dx^2_{4d} + d\rho^2
\end{equation}
and the solution preserves half of the original supersymmetries if the
supersymmetry rules (\ref{transfos1}) vanish for some Killing spinor
parameter $ \epsilon^i$.  Assuming that the domain--wall is supported
only by the scalars $q^\X(\rho)$, the flow equations become
\begin{equation}
\label{000}
\begin{array}{rcl}
\displaystyle U' &=& \pm g W ,\\[3mm]
q^{\X\prime} &=&  \mp 3 g \,g^{\X\Y} \partial_\Y W,
\end{array}
\end{equation}
and a solution of these equations is proved to solve the equations of
motion.


\section{Construction of non-homogeneous quaternionic spaces}


Our construction of 4--d quaternionic spaces that are not homogeneous
follows the procedure discussed by Page and Pope in \cite{060}.
The starting point is a 2--d space $\Sigma$ with constant positive,
negative or vanishing curvature given by the metric and volume (or
K\"ahler) form\footnote{In what follows,  when clear from the context,
we will mostly understand the tensorial or wedge products.}
\be124
d\Sigma_{\kappa} =  
{d\zeta \otimes d\bar \zeta \over \left(1 + {\kappa\over 4} 
\zeta \bar \zeta\right)^2} \quad ,\quad
K_{\kappa} ={d\zeta 
\wedge d\bar \zeta \over \left(1 + {\kappa\over 4} 
\zeta \bar \zeta\right)^2} 
\ee
where $\kappa$ is the constant curvature, i.e.\ $\kappa = 1$ for a sphere,
$\kappa=-1$ for a hyperboloid or for flat space $\kappa=0$. 
As next step one adds a non-trivial line bundle over this space and
writes the 4-d metric as
\be399
ds^2 = {dr^2 \over V(r)} + V(r) \left( d\tau + A \right)^2 + 
    F^2(r) \, d\Sigma_\kappa 
\ee
where we used the freedom to introduce a proper radial coordinate so
that we are dealing, in addition to the 1-form $A$, with only two unkown
functions $V(r)$ and $F(r)$.  These functions and the 1-form are fixed
by the requirement to have an Einstein metric solving the equation
\be993
R_{_{XY}}= - {3 \over l^2} \, g_{_{XY}} \ .
\ee
For the 1--form $A$ this means that ${dA}$ is proportional to the K\"ahler
form of the 2--d base space with the metric $d\Sigma_\kappa$ \cite{060} and
therefore
\be884
A = n \, {i \over 2} \, {\zeta d\bar \zeta - \bar \zeta d\zeta \over 
1 +{\kappa\over 4} \zeta \bar \zeta}
\ee
with some constant $n$. 
Note, for the compact case ($\kappa=1$) this 1--form is not globally
defined and it differs in each coordinate patch resulting in a
periodicity condition for $\tau$ which depends on $n$.
We will come back to this point below.  

The function $F(r)$ can be obtained from $R_r^{\ r} - R_{\tau}^{\
\tau}=0$ yielding the equation ($F' = {d F \over dr}$)
\be294
F^3 F'' + n^2 = 0
\ee
which is solved by
\be234
F^2 = c_1 (r+ c_2)^2 - {n^2 \over c_1} = c_1 r^2 + 2 c_1 c_2 r
    +{c_1^2 c_2^2 - n^2 \over c_1} \ .
\ee
If $c_1 \neq 0$, both integration constants can be absorbed
by rescalings of $r$, $n$, $V$, $\tau$ and this function becomes
\be320
F^2 = r^2 - n^2 \ .
\ee
On the other hand, the case $c_1 = 0$ requires that $c_1 c_2 = n$ and
the solution is $F^2 = 2 n \, r $, which is however equivalent
to the case before in the limit $n \rightarrow \infty$ while keeping
$\hat r = n (r -n)$ fixed; we come back to this limit below.  
Finally, in order to fix $V(r)$ we consider the equation $R_{r}^{\ r}
- R_x^{\ x} = 0$ ($x = {\rm Re}(\zeta)$) which, after inserting
$F(r)$ in the metric, yields the differential equation
\be227
(r^2 - n^2)^2 \, V'' - 2(r^2 + 3 n^2) \, V + 2\kappa \, (r^2 - n^2) = 0
\ee
and is solved by
\be662
V(r) = {1 \over r^2 - n^2} \left[ (r^2 + n^2) \kappa  - 2 m r + 
    {1 \over l^2} (r^4 -6 r^2 n^2 -3 n^4) \right] 
\ee 
where $m$ is an arbitrary integration constant and we identified the
second integration constant with the cosmological constant $l^2$.

For a special choice of parameters this solution includes the
AdS--Taub--NUT and AdS--Taub--Bolt solution as well as the coset space
$SU(2,1)/U(2)$, see \cite{book, 010}.  
But before we discuss special cases, let us determine whether the Weyl
tensor is anti-selfdual, which indicates that this 4-dimensional space
is quaternionic.
One finds that ${\cal W}^+ = {1 \over 2} ({\cal W} + {^{\star}{\cal W}} )=0$ 
only if the mass parameter is given by
\be224
m = n \kappa - 4 \, {n^3 \over l^2} \ .
\ee
Similarely, the anti-selfdual component vanishes ($W^- =0$) if 
$\displaystyle m = -n \kappa + 4 \, {n^3 \over l^2}$.  
If we want to consider this space as the moduli space of a single
hypermultiplet, we have in addition to ensure (\ref{Ricci}), 
that is $R_{_{XY}} = - 3 \, g_{_{XY}}$ or $l^2 = 1$, and finally obtain
\be235
\ba{rcl}
ds^2 &=& \displaystyle { dr^2 \over V} +V\left( d\tau + n {i\over 2} 
    { \zeta d \bar \zeta - \bar \zeta d \zeta \over 1 + 
    {\kappa\over 4} \zeta \bar \zeta}\right)^2  + (r^2 - n^2) \, 
    {d\zeta d\bar \zeta \over (1 + {\kappa\over 4} 
    \zeta \bar \zeta)^2} \ , \\[4mm]
V &=& \displaystyle {r - n \over r + n} \left[(r+n)^2 + 
    (\kappa - 4 n^2) \right] \ .
\ea
\ee
It is straightforward to verify that this metric satisfies the
relations discussed in the previous section.

This metric is not well-behaved everywhere, possible dangerous points
are zeros and poles of $V$, but also the point $|\zeta|^2 = - 4 /
\kappa$.
Most of these points are anyway coordinate artifacts, because the square of
the Riemann tensor reads
\be991
R_{_{XYZW}} R^{^{XYZW}} = 24 + 96 \, n^2 \, {(\kappa - 4 n^2)^2 
    \over (r+n)^6}
\ee
and hence only at the pole of $V$ at $r = -n$ we expect a curvature
singularity.

\bigskip

To show explicitly the quaternionic structure of the above space and 
its characteristic quantities, we introduce the following one--forms:
\begin{equation}
\begin{array}{rcl}  
u&=&\displaystyle \frac{1}{\sqrt{2}} \sqrt{r^2 - n^2} \frac{d\bar \zeta}
{1+\frac{\kappa}{4} \zeta \bar \zeta}\ ,
\\[4mm]
v&=& \displaystyle\frac{1}{\sqrt{2}} \left[\frac{dr}{\sqrt{V(r)}} + {\rm i} 
\sqrt{V(r)} \left(d\tau +  {i\over 2} n
{ \zeta d \bar \zeta - \bar \zeta d \zeta \over 1 + 
{\kappa\over 4} \zeta \bar \zeta}\right)\right]\,.
\end{array}
\end{equation}
The quaternionic vielbeine $f^{iA} = dq^\X \, f^{iA}_{\X}$ are
\begin{equation}
f^{iA} = \left(\begin{array}{cc} u & -v \\ \bar v & \bar u \end{array} 
\right),
\end{equation}
and the metric (\ref{235}) is given by $g = f^{iA} \otimes f_{iA} = 2
u \otimes \bar u + 2 v \otimes \bar v$.  From these vielbeine one
obtains for the $SU(2)$--connection and $SU(2)$ curvature
\begin{equation}
\omega^r =  -\sqrt{\frac{V(r)}{8}}\left( \begin{array}{c}
\displaystyle
{\rm i}\; \frac{ u - \bar u}{r- n}\\[4mm]
\displaystyle
\frac{ u+\bar u }{r- n}\\[4mm]
\displaystyle
{\rm i} \frac{r-n}{ V(r)} (  v - \bar v) + \frac{\rm i}{\sqrt{8V}} 
\kappa  { \zeta d \bar \zeta - \bar \zeta d \zeta \over 1 + 
{\kappa\over 4} \zeta \bar \zeta}
\end{array} \right)
\end{equation}
and
\begin{equation}
{\cal R}^r = \frac12 \left( \begin{array}{c} {\rm i} (uv + \bar u \bar v) \\
uv  - \bar u \bar v \\ - {\rm i}(u \bar u + v\bar v)
\end{array} \right)
\end{equation}
respectively.

\bigskip

It is interesting to note that this quaternionic space interpolates
between the two homogeneous spaces ${SO(4,1) \over SO(4)}$ and
$\frac{SU(2,1)}{U(2)}$. 
Let us discuss this in more detail.

\medskip
\pagebreak

\noindent 
{\em (i) $AdS$ limit: $n=0$}

\medskip

\noindent 
The metric of the 4-dimensional Euclidean anti-deSitter space can be
written as
\be229
ds^2 = {dr^2 \over \kappa + {r^2 \over l^2} } + 
    \Big(\kappa + {r^2\over l^2} \Big) d\tau^2 + r^2 
    d\Sigma_\kappa \ .
\ee
Obviously, the metric (\ref{235}) becomes exactly $AdS_4$ if $n=0$.

\medskip

\noindent 
{\em (ii) ${SU(2,1) \over U(2)}$ limit: $n \rightarrow \infty$}

\medskip

\noindent 
In order to have a regular large $n$ limit we replace first
\be881
r = {\hat r \over n} + n \qquad , \qquad \tau = 2n \psi
\ee
and keep $\hat r$ fixed while taking the limit $n \rightarrow
\infty$. 
As a result the metric becomes
\be913
ds^2 = {d \hat r^2 \over 2 \hat r (\hat r + {\kappa\over 4})}
    + 8 \hat r \, \Big( \hat r + {\kappa \over 4} \Big) \Big[
    d\psi + {i\over 4} { \zeta d \bar \zeta - \bar \zeta d \zeta
    \over 1 + {\kappa\over 4} \zeta \bar \zeta}\Big]^2
    + 2 \hat r \, d \Sigma_\kappa
\ee
which is equivalent to the $c_1 = 0$ case in (\ref{234}).  

For $\kappa = 1$ we can perform the change of coordinates 
\begin{equation}
\label{angul}
\hat r = {\rho^2 \over 4 ( 1- \rho^2)}, \quad 
\zeta = 2 \tan {\theta \over 2} \, e^{- i \varphi},
\end{equation}
which yields
\be118
ds^2 = {2\, d\rho^2 \over (1- \rho^2)^2} + { \rho^2 \over 2(1-\rho^2)^2}
    \left[ d\psi + (\cos\theta - 1 ) d\varphi \right]^2 + {\rho^2 \over
    2 (1-\rho^2)} \left[d\theta^2 + \sin^2\theta d\varphi^2 \right].
\ee
This is (up to an overall factor of ``2'' to ensure the correct value
of the Ricci scalar) the known Bergman metric giving a 
parameterization of the coset ${SU(2,1) \over U(2)}$.  

For $\kappa = 0$ we get another parameterization of this coset, which
appears naturally in string theory compactification.
The corresponding transformation reads $2\psi = {1 \over 2i} (S - \bar
S)$, $2 \hat r = e^{2 \phi} = ( {S + \bar S \over 2} - \zeta \bar
\zeta )^{-1}$ and brings the metric in the form
\be711
ds^2 = 2 \, e^{4 \phi} \Big( {1 \over 2} dS - \bar \zeta d \zeta \Big)
    \Big( {1 \over 2} d\bar S - \zeta d \bar \zeta \Big) + 2\, e^{2\phi}
    d \zeta d\bar \zeta
\ee
which is known to describe the geometry of the classical moduli space
of the universal hypermultiplet \cite{160}.  Hence, while varying the
parameter $n$ we can smoothly interpolate between the two homogeneous
spaces.

To get a better understanding of this parameter, let us discuss the
case $\kappa=1$ in more detail. 
In this case the metric can be written as
\be994
ds^2 =  {dr^2 \over V(r)} + V(r) \, 
    \left[ d\tau + 2n \, \cos\theta \, d\varphi\right]^2
    + (r^2 - n^2) \left[ d\theta ^2 + \sin^2 \theta \, d\varphi^2 \right]
\ee
but this coordinate system that we are using here is globally not well
defined.  In fact, in order to avoid a Dirac singularity one has to
introduce different 1--forms $A$ on the north- and south hemisphere of
the $S_2$ (resulting in the replacement $\cos\theta \, d\varphi
\rightarrow ( \cos\theta \mp 1) \, d\varphi$) and in order to ensure a
smooth interpolation at the equator, the periodicity of $\tau$ has to
be related to $n$ by
\be773
\tau \simeq \tau + 8 \pi n \ .
\ee
This changing of the periodicity can also be understood as a result of
an orbifold acting on the Euclidean ${\mathbb E}AdS_4$ space; for
orbifolds acting on Minkowskean $AdS$ spaces we refer to \cite{030,
180, 170} and references therein.  One may have expected this
interpretation, since in the limit of vanishing cosmological constant
one gets the Taub--NUT metric \cite{140}, which is a resolution of the
${\mathbb R}_4/{\mathbb Z}_n$ orbifold and for a non--zero
cosmological constant ${\mathbb R}_4$ is replaced by an Euclidean
${\mathbb E}AdS_4$ space.  By complete analogy to the Taub--NUT case
we can make the orbifold action explicit.  Namely, 
${\mathbb E} AdS_4 = {SO(4,1) \over SO(4)}$ is defined by the
hyperboloid
\be263
- (X_0)^2 + (X_1)^2 + (X_2)^2 + (X_3)^2 + (X_4)^2 = - l^2 \ .
\ee
The metric can be obtained by starting with the flat space metric
\be274
ds^2 = - (dX_0)^2 +(dX_1)^2 + (dX_2)^2 + (dX_3)^2 + (dX_4)^2 
\ee
and imposing the constraint (\ref{263}).  
Before imposing the constraint, the $SO(4,1)$ symmetry group is
manifest, but afterwards only a subclass of these isometries are
realized linearly and the other symmetries are not manifest.
We are interested here in the spherical symmetric case ($\kappa =1$) and an
obvious way to keep this symmetry manifest is by introducing polar
coordinates in $X_{1,2,3,4}$: $(dX_1)^2 + (dX_2)^2 + (dX_3)^2 +
(dX_4)^2 = d \rho^2 + \rho^2\, d \Omega_3$ and the constraint becomes
$X_0^2 - \rho^2 = l^2$.
Since $X_0 \, dX_0 = \rho \,d \rho$ we can eliminate the timelike
coordinate and find for the metric
\be188
ds^2 = {d\rho^2 \over 1+ {\rho^2 \over l^2}} + \rho^2 d \Omega_3
    = {d \bar r^2 + \bar r^2 d\Omega_3 \over 
    (1 - {\bar r^2 \over 4 l^2})^2}
    = {dz_1 d\bar z_1 + dz_2 d \bar z_2 \over 
    \left(1 - {|z_1|^2 + |z_2|^2 \over 4 l^2} \right)^2}
\ee
with $\rho = { 2 l^2 \, \bar r \over l^2- \bar r^2}$.  As for the
Taub-NUT space the ${\mathbb Z}_n$ orbifold acts on the two complex
coordinates as the following identification
\be774
z_1 \simeq e^{2\pi i \over n} z_1 \qquad , \qquad z_2 \simeq 
    e^{-{2\pi i \over n}} z_2 
\ee
which, after the change of coordinates
\be385
z_1 = r \, \cos(\theta / 2) \, e^{i {\varphi + \psi \over 2}}
\, , \quad 
z_2 = r \, \sin({\theta / 2}) \, e^{i {\varphi - \psi \over 2}}\, ,
\ee
corresponds to
\be663
\psi \simeq \psi +{4\pi \over n} \ .
\ee
Therefore the $S_3$ of the ${\mathbb E}AdS_4$ space is replaced by the
well-known Lens space $S_3/Z_n$.  This identification breaks however
part of the isometry group, which becomes clear if we write the
orbifold as
\be362
 \left( \ba{c} z_1 \\ z_2 \ea \right) \simeq
     \left( \ba{cc} \lambda & 0 \\ 0 & {1 \over \lambda}  \ea \right)\,
     \left( \ba{c} z_1 \\ z_2 \ea \right) = \Gamma 
     \left( \ba{c} z_1 \\ z_2 \ea \right) 
\ee
with the complex phase $\lambda = e^{2\pi i \over n}$.  
The $SO(4,1)$ isometry group has the maximal compact subgroup $SO(4)
\simeq SU(2)\times SU(2)$.
One of the $SU(2)$ rotates the two complex coordinates and obviously
only its diagonal subgroup commutes with the orbifold action $\Gamma$.
The other $SU(2)$ does not mix the two complex coordinates and
commutes with the orbifold.
The isometry group contains moreover four Lorentz boosts, which also
do not commute with $\Gamma$, and hence the orbifold breaks the
isometry group down to $U(1) \times SU(2)$.

In supergravity one imposes the orbifold in the asymptotic space by
introducing a conical deficit angle by a modified metric ansatz.  
In the case at hand one replaces the $S_3$ metric 
\be364
d\Omega_3 = \left[ d\tau + \cos\theta \, 
    d\varphi\right]^2 + d\theta^2 + \sin \theta^2 d\varphi^2 \ .
\ee
by
\be824
\left[ d\tau + 2n \, \cos\theta \, 
    d\varphi\right]^2 + d\theta^2 + \sin^2 \theta\,  d\varphi^2 
\ee
where $n$ is the NUT charge and as discussed before this is consistent
as long as $\tau$ has the periodicity $8 \pi n$.

Thus, the orbifold effectively changes the periodicity of the compact
$\tau$ and fixed points of the orbifolds are given by fixed points of
the Killing vector $k = \partial_{\tau}$.  If we write
\be119
V(r) = {r -n \over r+n} \, \left(r - r_-\right)\left(r - r_+\right) 
    \quad , \qquad
    r_{\pm} = -n \pm \sqrt{4 n^2 - \kappa}  
\ee
the fixed points are at $r = n$ and at $r = r_{\pm}$.  Note, there is
no need to assume that $r$ is positive -- any coordinate region in
which the metric is Euclidean is an allowed physical parameter range.
This is the case as long as $V(r) \geq 0$ and $r^2 \geq n^2$.
Therefore, for each $\kappa$ we have two (disconnected) fixed points:
at $r = r_-$ (with $r \leq r_-$) and at $r = n$ for $\kappa \geq 0$
(with $r \geq n$) or at $r = r_+$ for $\kappa = -1$ (with $r \geq
r_+$).

As it is obvious from curvature square, see eq.\ (\ref{991}), all
these points are regular points of the manifold, but the different
models imply a different periodicity of $\tau$ (fixed by the absence
of conical singularities).  Consider the $(r,\tau)$--part of the
metric in the neighborhood of these points.  We start with the case
$\kappa=1$ which corresponds to the orbifold discussed before and
consider $\rho = r-n \simeq 0$.  Then $V \simeq {\rho \over 2n}$
and the $(r,\tau)$--part of the metric becomes (with $\bar r^2 = 8n
\rho$)
\be912
ds^2 = {2n \, d\rho^2 \over \rho} + {\rho\over 2n } d\tau^2
    = d\bar r^2 + {\bar r^2 \over (4 n)^2} \; d\tau^2
\ee
which has no conical singularities as long as $\Delta \tau = 8 n \pi$
in accordance to the identification given in (\ref{773}). 
Next, set $\kappa=0$.
Now $r= n$ becomes a double zero of $V(r)
\simeq 2 \rho^2$ and the $(r,\tau)$--part of the metric reads
\be669
ds^2 = {d\rho^2 \over 2 \rho^2} + 2 \rho^2 d\tau^2 \ .
\ee
This is an $AdS_2$ metric implying no periodicity in $\tau$. 
For $\kappa=-1$ the fixed point is at $r=r_+$ ($r=n$ is outside the
allowed physical region) and expanding around $\rho = r-r_+ \simeq 0$
we find for the $(r,\tau)$--part of the metric ($\rho = {\beta_+ \bar
r^2 \over 4}$)
\be551
ds^2 = {d\rho^2 \over \beta_+ \rho} + {\beta_+ \rho } \, d\tau^2
    = d\bar r^2 + \beta^2_+ \, \bar r^2 \; d\tau^2
\ee
with $\beta_+ = 2(r_+ - n) $ and hence, the periodicity of $\tau$
becomes $\tau \simeq \tau + {2 \pi \over \beta_+}$. 
Finally, as we mentioned before there is a second allowed coordinate
range ($r \leq r_-$) with the fixed point $r = r_-$.
In analogy to the previous cases one obtains the following
periodicity: $\tau \simeq \tau + {2 \pi \over \beta_-}$ with
$\beta_- = 2 (n- r_-) $.

The different periodicities in $\tau$ may sound inconsistent, but for
each model given by a value of $\kappa$ and defined in a physical
parameter range there is exactly one periodicity of $\tau$ which
avoids a conical singularity at the accessible fixed point.


\section{Isometries and their gauging}


In this section we would like to use the non--homogeneous scalar
manifolds constructed in the previous ones as moduli space of one
hypermultiplet coupled to five--dimensional supergravity theory.  In
particular, we are interested in trying to understand the possibility
of obtaining supersymmetric vacua from the scalar potential resulting
from gauging a linear combination of their isometries.  To do so, we
first describe the isometry structure emerging from the definition of
the metric of such manifolds.  We then show how the algebraic nature
can be understood in terms of embeddings into the five--dimensional
Lorentz group and finally discuss the critical points.

To simplify the analysis of the isometries and the gauging of their
linear combinations, in the following we choose to parameterize the
quaternionic scalar manifold by the coordinates $q^X = \{ r,\tau,x,y\}$
with the metric given by (\ref{235}), with $\zeta = 2(x + {\rm i} y)$,
\begin{equation}     \label{metr60}
\begin{array}{rcl}
ds^2 &=& \displaystyle  \frac{dr^2}{V(r)}+ V(r) \left[d\tau + 4\, 
 n \, \frac{x \, dy - y\, dx}{\left[1 + \kappa (x^2 + 
y^2)\right]}\right]^2 +\\[5mm]
&+&\displaystyle 4 \frac{r^2 - n^2}{\left[1 + \kappa (x^2 + 
y^2)\right]^2} \left(dx^2 + dy^2\right) \ .
\end{array}
\end{equation}
It was shown in \cite{book} that these manifolds have four isometries
and using the above basis they are given by
\begin{equation}
\label{kill1}
\begin{array}{rclrcl}
k_1 &=& \left(\begin{array}{c} 0\\ 1\\ 0\\ 0\end{array} \right),&
k_3 &=& \left(\begin{array}{c} 0\\ 4\,n \;x \\ 2\kappa \;x y \\
      1 - \kappa \,( x^2 -y^2)\end{array} \right),\\[10mm]
k_2 &=& \left(\begin{array}{c} 0\\ 0\\ -y\\ x\end{array} \right),&
k_4 &=& \left(\begin{array}{c} 0\\ -4\,n \;y \\
      1 + \kappa  \,( x^2 -y^2)\\ 2\,\kappa\; x y\end{array} \right).
 \end{array} 
\end{equation} 
Using (\ref{valuePrI}) we can also
compute the value of their prepotentials:
\begin{equation}
    \label{123}
\begin{array}{rcl} 
P_1 &=& \left( \begin{array}{c} 0 \\ 0 \\ \displaystyle\frac14 
(r-n)\end{array} \right)\,, \\[10mm]
P_2 &=& \displaystyle\frac{1}{4
\left(1+ \kappa (x^2 + 
y^2)\right)}\left( \begin{array}{c} -2x \sqrt{(r-r_-)(r-r_+)} \\[1mm]  
2y \sqrt{(r-r_-)(r-r_+)}  \\[1mm]  -1+(x^2 + 
y^2)(\kappa+ 4nr-4n^2)\end{array} \right)\,, \\[10mm]
P_3 &=&\displaystyle\frac{1}{\left(1+ \kappa (x^2 + 
y^2)\right)}\left( \begin{array}{c} \frac{-1+\kappa(x^2 - y^2)}{2}
\sqrt{(r-r_-)(r-r_+)} \\[1mm] 
-\kappa \, x y \sqrt{(r-r_-)(r-r_+)}  \\[1mm]  x (\kappa - 2 n^2 + 2 
n r) \end{array} \right)\,,\\[10mm]
P_4 &=& \displaystyle\frac{1}{\left(1+ \kappa (x^2 + 
y^2)\right)}\left( \begin{array}{c} 
-\kappa \, x y \sqrt{(r-r_-)(r-r_+)}  \\[1mm]  
-\frac{1+\kappa(x^2 - y^2)}{2}
\sqrt{(r-r_-)(r-r_+)} \\[1mm]  
-y (\kappa - 2 n^2 + 2 n r) \end{array} \right)\,.
\end{array}
\end{equation}

To understand the algebraic structure generated by these four 
isometries, one  can work out their commutation relations.
The result is that $k_1$ commutes with all the other Killing vectors
\begin{equation}
[k_1 , k_{i}] = 0, \quad i \in 1,\ldots,4
\end{equation}
whereas the others satisfy the following relations:  
\begin{eqnarray}
\, [k_2,k_3] &=&  k_4\ ,\\
\, [k_4,k_2] &=&  k_3\ ,\\
\, [k_3,k_4] &=&  4 \kappa \, k_2 - 8 n \, k_1\ .
\end{eqnarray}

The structure of the algebra formed by such Killing vectors becomes
more transparent if one introduces for $\kappa \neq 0$ the following
combinations
\begin{equation}
    \label{Tgen}
T_1 \equiv \frac12 k_3, \quad T_2 \equiv \frac12 k_4, \quad T_3 
\equiv \kappa k_2 - 2 n k_1 \ \  \hbox{ and } \ T_4 \equiv k_1 \ .
\end{equation}
Their commutation relations are given by
\begin{equation}
\begin{array}{rcl} 
\, [T_i,T_j] &=&  f_{ijk} T_k, \quad i,j,k\in \{1,2,3\},\\[3mm]
\, [T_4,T_i] &=&  0,
\end{array}
\end{equation}
with $f_{123} = 1$, $f_{231} = \kappa$ and $f_{312} = \kappa$.  One
can now easily see that for $\kappa = 1$ the group of the isometries
is $SU(2) \times U(1)$, with $f_{ijk} = \epsilon_{ijk}$ whereas for
$\kappa = -1$ it becomes $SL(2, {\mathbb R})\times U(1)$ (recalling
that $SL(2, {\mathbb R})\simeq SO(2,1)$ ).  A bit more elaborated is
the case when $\kappa = 0$.  In this case indeed the group generated
by the four isometries is no more the product of simple groups, but we
obtain a solvable group whose algebra contains the Heisenberg algebra.
This latter is realized by the $ k_1, k_3 $ and $k_4$ generators.

It seems interesting to point out that the three cases share the
common feature of being different realizations of subgroups of the
five--dimensional Lorentz group $SO(1,4)$.  It is also worth noting
that this same group describes the isometries of the Euclidean version
of the four--dimensional $AdS$ space, which is given by the ${\mathbb
E}AdS_4 =SO(1,4)/SO(4)$ coset.  This fact was indeed expected for the
$\kappa = 1$ case.  Indeed, as it was shown in the previous section,
this manifold can be obtain as an orbifold of the ${\mathbb E}AdS_4$
space.  Moreover one can show that its isometries are those which
survive this orbifold action, i.e. those which commute with the
generator of the orbifold.  In detail, we can define the standard
generators of the $SO(1,4)$ group in the adjoint representation
$(T^{ab})$, such that they satisfy the usual commutation relations
\begin{equation}
\ [ T^{ab}, T_{cd}] = 4 \delta^{[a}{}_{[c} T^{b]}{}_{d]} \ ,
\end{equation}
with the indices raised and lowered by the $SO(1,4)$ metric 
$\eta_{ab}=\hbox{diag}\{-++++\}$.
The compact subgroup is then generated by  $T^{\hat a \hat b}$ with 
$\hat a  = 1,2,3,4$.
If we now choose for this basis the adjoint representation given by $
(T^{ab})_{cd} = \delta^{ab}_{cd}$, the orbifold
action (\ref{362}) corresponds to the discrete element given by
\begin{equation}
T_{orb} = e^{ \frac{2\pi {\rm i}}{n} (T^{12} - T^{34})}.
\end{equation}
In the same way, one can see that the group of the isometries of the 
$\kappa = -1$ manifold  is given in terms of the 
subgroup of the $SO(1,4)$ group which commutes with one of the 
compact generators which sit in one of the two $SU(2)$ groups of the 
$SO(4) \simeq SU(2) \times SU(2)$ decomposition, for instance
\begin{equation}
e^{\frac{2\pi {\rm i}}{n} T^{12}}.
\end{equation}
As its structure is not given by the product of simple groups, 
the case $\kappa = 0$ cannot be obtained in the same way, but it is 
still a solvable sub-algebra of the $SO(1,4)$ algebra.

\begin{figure}[htbp]
    \centering
\includegraphics[angle=0,width=50mm]{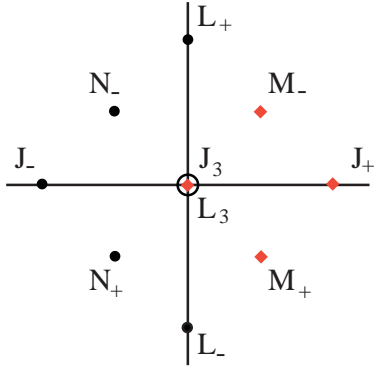}
    \caption{Root diagram for the $SO(1,4)$ group.}
    \label{figura}
\end{figure}

An easy way to understand these results without having to compute 
all the commutation relations of the generators can be provided by 
the analysis of the root  diagram of $SO(1,4)$ as presented in 
picture (\ref{figura}).

The roots are the various generators, displayed according to their
weight, measured by their commutator with the two Cartan generators
$ J_3 =  (T^{12} + T^{34})$ and $L_3 = (T^{12} - T^{34})$.
The other compact generators are $J_{\pm}$ and $L_{\pm}$, which 
complete the $SU(2)\times SU(2) \simeq SO(4)$ stability group.
The non--compact generators (boosts) are $M_{\pm}$ and $N_{\pm}$.

In these conventions, the $T_{orb}$ generator is given by either
$e^{\frac{2\pi {\rm i}}{n} J_3}$ or $e^{\frac{2\pi {\rm i}}{n} L_3}$
and therefore the elements commuting with it are all the generators on
the line orthogonal to the $(1,0)$ or $(0,1)$ vector in the root
space.  This implies that the remaining group generators are the sets
$\{ J_\pm, J_3, L_3\}$ or $\{L_\pm, J_3, L_3\}$ which span
$SU(2)\times U(1)$.  On the other hand, if we now look at the $\kappa
= -1$ case, the generator which defines the surviving isometry group
can be chosen to be $e^{\frac{2\pi {\rm i}}{n} (J_3\pm L_3)}$.  The
residual symmetry group is then generated by the elements commuting
with it and these are all the generators on the line orthogonal to the
$(1,\pm 1)$ vector.  The resulting set will be then given by $\{ N_-,
M_+, J_3 \pm L_3\}$ or $\{ N_+, M_-, J_3 \pm L_3\}$, both generating
$SL(2,{\mathbb R})\times U(1)$.  As already remarked, the case $\kappa
= 0$ cannot be obtained in this same way.  One can anyway see that the
correct commutation relations are obtained by any solvable sub-algebra
of the $so(1,4)$ one which sits on one of the corners of the diagram.
One example is given by the red diamond generators in Figure
(\ref{figura}), namely $\{ J_3, L_3, J_+, M_{\pm}\}$.

\bigskip

The ${\mathbb E}AdS_4$ space is a symmetric quaternionic space and
therefore the analysis of the supersymmetric vacua obtained by gauging
its isometries easily follows from the considerations made in
\cite{flatdw}.  Indeed it has been shown that if one looks only at the
theory of supergravity coupled with hypermultiplets, the isometries
leading to critical points are identified as those lying in the
stability subgroup of the isometry group.  The result is that one can
obtain supersymmetric critical points of the potential by gauging
isometries in $SO(4)$.

We have just seen that the isometries of our non--homogeneous
quaternionic spaces can be understood as subsets of those of such
${\mathbb E}AdS_4$ space and it can also be shown that in the limit
which gives back such space $n\to 0$, the isometries can indeed be
identified.  As a consequence, one could expect that the possibility
of obtaining supersymmetric vacua from the gauging of the
(\ref{kill1}) isometries depends on such identification.  For
instance, from the above analysis, a critical point should follow from
the gauging of any isometry in the $\kappa = 1$ case, but only for
special combinations in the $\kappa = -1$ and $\kappa = 0$ cases.


\section{Critical points and a domain wall solution}


As next step we will give all critical points that can be obtained by
the gauging described before. We will start with some general remarks
followed by a detailed analysis of the critical points for the given
model. A novel feature of this model is that it allows for multiple
critical points such that we can construct a smooth domain
wall solution interpolating between them.


\subsection{Characterizing critical points}


For a given superpotential it is a non-trivial question to determine
all critical points, but in gauged supergravity, critical points can
be directly determined from the Killing vectors \cite{130, 260, 450,
flatdw}.  In our case, where only a single isometry has been gauged
the situation is especially clear.

A direct consequence of the flow equations, is that the supersymmetric
flow goes perpendicular to the Killing vector $k$ and hence terminates
at points where $k$ becomes null, i.e.\ $|k|^2 = 0$.  This fixed point
set spans an even-dimensional submanifold.  To see this one considers
a parallel transport of the Killing vector $k$ along a vector $l$ so
that $\delta k_{\X} \sim (\nabla_{\Y} k_{\X}) l^{\Y} = (\partial_{[\Y}
k_{\X]} ) l^{^Y}$; where we assumed a regular coordinate system and
used the Killing property.  Hence, if $dk \neq 0$ the number of
possible translations along the null hypersurface is given by the rank
of the 2-form $dk$ calculated on the null surface.  If there are no
non-trivial translations, the null hypersurface is just a point of the
manifold, a so-called nut.  If the rank is two implies that there are
two translations that leave the null hypersurface and the other two
stay inside.  Hence the corresponding fixed point set is a
2-dimensional surface or a so-called bolt, see \cite{080} for more
details.  In both cases one can construct a flow that terminates at
the fixed point where $|k|^2 =0$.  In the degenerate case however,
i.e.\ if $(\nabla_{\X} k_{\Y})( \nabla^{\X} k^{\Y}) = 0$, the flow can
never reach the null hypersurface.  To see this note that in the
degenerate case the Killing vector {\em and} all derivatives vanish on
the hypersurface; recall any Killing vector obeys the relation
$\nabla_{\X}\nabla_{\Y} k_{\Z} = R_{\Z\Y\X\W} k^{\W}$.  The Killing
vector is therefore non--analytic and there is no possible
perturbative expansion around such a point.  Since the Killing vector
is non-trivial in the interior we infer that a flow can never reach
this point and exhibits a run-away behavior.  In addition, in the
degenerate case the surface gravity vanishes, whereas for the
non-degenerate case not.  As it is known from black holes, a
non-vanishing surface gravity implies a periodicity of the coordinate
along the Killing direction, whereas in the degenerate case this
coordinate is non-compact.  This is in agreement with the situation
for homogeneous spaces where a gauging of a compact direction yields a
fixed point, which may result in a flat or anti-deSitter spacetime
whereas the gauging of a non-compact directions produces a run-away
solution \cite{flatdw}.  Let us add a further side remark.  Especially
in the case of a 4-dimensional quaternionic space, one can express the
superpotential only by the discussed 2-form $dk = \partial_{\X} k_{\Y}
dq^{\X} \wedge dq^{^Y}$.  Namely, for flat tangent space indices we
can use the relation for our complex structures
\be829
\sum_r J^r_{\X\Y} J^r_{\Z\W} = - \epsilon_{\X\Y\Z\W} +
    \big(\delta_{\X\Z} \delta_{\Y\W} -\delta_{\X\W} \delta_{\Y\Z} 
    \big)
\ee
and write with (\ref{Wmydef}) and (\ref{valuePrI}) for the
superpotential
\be889
W^2 = {1 \over 3} \Big( dk\wedge {^{\star} dk} - {1 \over 2} 
    dk \wedge dk \Big) \ .
\ee
For higher dimensional quaternionic spaces an analogous relation holds,
which however includes curvature terms.

In summary, good critical points resulting in vacua with fixed
or constant scalars, are given if the following relations
are satisfied
\be619
|k|^2 = g_{\X\Y} k^{\X} k^{\Y} = 0 \quad , \quad
(\nabla_{\X} k_{\Y})(\nabla^{\X} k^{\Y}) \neq 0 \quad , \quad
R_{\X\Y\Z\W} R^{\X\Y\Z\W} \neq \infty \ .
\ee
These relations are independent of the choosen coordinate system, but
of course at these points one can always introduce a regular
coordinate system and can infer that the Killing vector itself has to
vanish \cite{130, 450, 260, flatdw}, which gives simpler equations.
But note, in most cases there are no good global coordinates and it is
better to use a coordinate independent notation.

For our model the situation is not so involved.  The only delicate
points in the metric are the zeros and the pole of $V$.  The pole at
$r = -n$, see eq.\ (\ref{991}), represents a curvature singularity and
we have to exclude this point.  On the other hand the zeros of $V$ are
regular points of the manifold and hence there is a regular coordinate
system so that the metric is smooth there.  As an example let us
consider the Killing vector $k_1 = \partial_{\tau}$.  In this simple
case $|k|^2 = V =0$ at $r=r_{\pm},n$.  At  $r=r_{\pm}$ the fixed point
set represents a bolt, because the 2-d base space $d\Sigma_{\kappa}$
remains finite, see (\ref{399}), and at $r=n$ we have a nut, where the
2-d base space together with the $U(1)$ fiber vanishes.  The surface
gravity of the nut is given by $(\nabla_{\X} k_{\Y})(\nabla^{\X}
k^{\Y}) \sim {\kappa^2 \over 4 n^2}$ and hence for $\kappa =0$ the nut
degenerates and is not a fixed point of the flow.  The other fixed
points at $r = r_{\pm}$ are always non-degenerate and the metric
becomes regular if we introduce the coordinates
\be667
z = {2 \over \sqrt{\beta_{\pm}}} \sqrt{r- r_{_\pm}} \, 
    e^{i \beta_{\pm} \tau} \ .
\ee
In fact as it clear from (\ref{551}) the $(r, \tau)$-part of the
metric becomes flat near the point $r = r_{\pm}$ and the Killing
direction $\tau$ is compact.
Moreover, after this transformation the new Killing vector becomes
\be883
k = \partial_{\tau} = \Big({d z \over d \tau} \, \partial_z + 
    {d \bar z \over d \tau} \, \partial_{\bar z} \Big) =
    i \beta_{\pm} \, (z \partial_z - \bar z \partial_{\bar z})
\ee
and the fixed point $r = r_{\pm}$ translates into $z = \bar z =0$,
which is a zero of the new Killing vector.


\subsection{Critical points}


Let us now consider the isometry obtained by the generic linear 
combination 
\be771
k = a^I k_I
\ee
and look for its zeroes.

The easiest case to analyze is $\kappa = 0$.  Critical points are
obtained only if
\begin{equation}
a_2 \neq 0 \; \hbox{ and } \;a_1 = 4 n 
\frac{a_3^2 + a_4^2}{a_2}
\end{equation}
and they lie at 
\begin{equation}
x = -\frac{a_3}{a_2}\,, \quad y = \frac{a_4}{a_2}\,.
\end{equation}
This means that for any value of the gauging parameter leading to 
critical points, this set is given by a plane in $r$ and $\tau$ for $x$ and 
$y$ fixed.
No further critical points are obtained at $r = n$, as can be understood 
from the analysis of $k^2$ and $(\nabla k)^2$.
The value of the cosmological constant at these 
critical points  is  given by
\begin{equation}
{\cal V}  = -\frac{a_2^2}{4}.
\end{equation}

\bigskip

For $\kappa = 1$, the analysis simplifies if one takes a different
definition for the gauged isometry.  Using the definition (\ref{Tgen})
of the $SU(2)\times U(1)$ generators given above, we can consider the
critical points of the generic isometry defined by
\begin{equation}
k = \sum_{r=1}^3 \alpha^r T_r + \beta T_4 \ .
\end{equation}

Let us now discuss first the case $r > n$ or $r< r_-$.
In this case, the critical points appear (i.e. $k^\X = 0$) anytime 
\begin{equation}
    \label{constr}
\beta = \mp 2 n ||\alpha||,
\end{equation}
with $||\alpha||^2 \equiv \sum_r \alpha_r^2$ being the $SU(2)$ norm of the 
parameters.
They sit at
\begin{equation}
\begin{array}{lcr}    
x &=& \alpha_1 \displaystyle \frac{\alpha_3 \pm ||\alpha ||}{(\alpha_1^2 + 
\alpha_2{}^2 )}\ , \\
y &=& \displaystyle -\alpha_2 \frac{\alpha_3 \pm ||\alpha ||}{(\alpha_1^2 + 
\alpha_2^2)}\ .
\end{array}
\end{equation}
Of course the result should be the same for any direction inside the
$SU(2)$ factor of the isometry and indeed, the limiting case $x=y=0$
for $\alpha_1 = \alpha_2 = 0$ and $\beta = 2 n \alpha_3$ 
should also be included\footnote{Taking the proper limit for the case 
$\beta = -2 n \alpha_3$ gives the points at $x \to \infty$ or $y \to 
\infty$, which are not covered by our stereographic parameterization
but which are nevertheless part of the manifold}.

This set is given by a plane parameterized by $r$ and $\tau$ for any
proper value of the gauging parameters.  The interesting feature to
note is that we cannot obtain critical points for any value of
$\alpha_r$ and $\beta$, but we have to satisfy the constraint
(\ref{constr}).  This leads to a mismatch with the expectations coming
from the previous group--theory analysis as we expected to have
critical points for any value of the gauging parameters.  This can be
solved by considering the points at the boundaries of our space $r=n$
and $r = r_-$, which are the points where the metric (\ref{metr60}) is
not well behaved.  Since we know that they are good points of the
manifold, we should extend our analysis to these points.  To do so one
can compute the $|k|^2$ invariant and see whether at such points it
vanishes.  If we take $r = n$, this is indeed the case for any of the
above isometries and $(\nabla k)^2$ is always different from zero at
such point.  This means that one always obtains at least one critical
point for any gauging of the $SU(2)\times U(1)$ isometries, as
expected.  Moreover, in complete analogy with what happens for
instance for the universal hypermultiplet \cite{flatdw}, certain
special combinations of two commuting $U(1)$'s make it degenerate to a
full plane.

In the $r \geq n$ region we have then found an $SU(2) \times U(1)$
symmetric critical point (in some cases connected to a plane of
critical points which preserve a $U(1)$).  This reflects into the
value of the cosmological constant at such critical points, which is
\begin{equation}
{\cal V}  = -\frac{|| \alpha ||^2}{4}
\end{equation}
and shows the $SU(2) \times U(1)$ invariance explicitly.  Again in
complete analogy with the homogeneous case, the cosmological constant
is generated by gauging the $SU(2)$ factor of the isometry, which will
then be interpreted as the $R$--symmetry of the theory.

To complete such analysis we now have to consider the points at $r =
r_-$ or $z=\bar z = 0$ in the regular parameterization of (\ref{667}).
Here the result is somehow surprising and quite different from all the
previous cases.  Looking at the zeroes of $k^2$, which are not zeroes
of $(\nabla k)^2$, one sees that {\it two distinct} critical points
appear for {\it any} gauging of the $SU(2)\times U(1)$ isometries:
\begin{equation}
\begin{array}{rcl}   
r &=& r_- \equiv  - n - \sqrt{4n^2 -1}\,,\\[2mm]
x &=& x_{\pm} \equiv \alpha_1 \displaystyle 
\frac{\alpha_3 \pm ||\alpha ||}{(\alpha_1^2 + 
\alpha_2{}^2 )}\,, \\[4mm]
y &=& y_\pm \equiv \displaystyle -
\alpha_2 \frac{\alpha_3 \pm ||\alpha ||}{(\alpha_1^2 + 
\alpha_2^2)}\,.
\end{array}
\end{equation}
As for the point $r = n$ in the previous case
we don't need to  impose (\ref{constr}).
These critical points preserve a $U(1) \times U(1)$ symmetry which is 
generated by the following combinations of the Killing vectors
\begin{equation}
    \label{kAkB}
\begin{array}{rcl}
k_A &=& \sqrt{4 n^2 -1} \, \alpha^r T_r + \beta \, T_4\\[2mm]
k_B &=& \sqrt{4 n^2 -1} \, \alpha^r T_r - \beta \, T_4
\end{array} 
\end{equation}
This again shows up in the value of the cosmological constant at the
critical points, which are given by
\begin{equation}
    \label{V+V-}
{\cal V}_{x_+,y_+} = -\frac14 (n- r_-)^2 A_+^2 \ , 
\quad \hbox{and} 
\quad {\cal V}_{x_-,y_-} = -\frac14 (n- r_-)^2 A_-^2\ , 
\end{equation}
for $A_\pm \equiv \beta \pm \sqrt{4 n^2 -1} ||\alpha||$.

A comment is in order here\footnote{We thank A. Van Proeyen for a
discussion on this issue.}.  As pointed out in \cite{math}, for a
large class of quaternionic--Ka\"hler manifolds, critical points which
are in the same branch of the moduli space must be connected through a
geodesic line of critical points.  It would be natural to expect the
same to happen here.  On the other hand we have shown above that these
two critical points are isolated.  The resolution of this apparent
paradox is given by the fact that our space is not geodesically
complete and that moreover the region containing the singularity is
excluded from physical requirement.  This implies that if we look for
the points where the Killing vectors are vanishing in the whole space,
we will see that indeed the above critical points are connected
through a geodesic critical line, but that this latter lies completely
inside the region of our moduli space which is not accessible.  This
results in two isolated points in the physical region, which are the
boundaries of such geodesic.  Moreover, as we will see in the next
section, we are now allowed to find a supersymmetric flow
interpolating between them.  Actually, this flow is driven by the
gradient of the superpotential and therefore does not necessarily
follow geodesic lines.

In the case $\kappa = -1$ the $SO(2,1)$ norm
\begin{equation}
||\alpha||^2 \equiv - \alpha_1 ^2 - \alpha_2^2 + \alpha_3^2
\end{equation}
can be both positive or negative.
If we analyze first the regions $r> r_+$ or $r<r_-$,  
we obtain critical points whenever $|| \alpha ||^2 \geq 0$ and
\begin{equation}
\beta = \mp 2 n || \alpha ||\ .
\end{equation}
The critical points sit at
\begin{equation}
\begin{array}{lcr}    
x &=& \alpha_1 \displaystyle \frac{\alpha_3 \pm ||\alpha ||}{(\alpha_1^2 + 
\alpha_2{}^2 )}\,, \\
y &=& \displaystyle -\alpha_2 \frac{\alpha_3 \pm ||\alpha ||}{(\alpha_1^2 + 
\alpha_2^2)}\,,
\end{array}
\end{equation}
exactly matching the $\kappa = 1$ case.
This similarity extends to the value of the cosmological constant, 
which is given by
\begin{equation}
{\cal V}  =  -\frac{|| \alpha ||^2}{4} \ .
\end{equation}
If, on the other hand $||\alpha||^2 < 0$, critical points can only 
appear for $x^2 + y^2 = 1$, which is out of the domain covered by our 
coordinate patch, cp.\ the metric in (\ref{metr60}).

Extending the analysis to the points $r = r_\pm$, for each of the
sections we find again two distinct critical points at $x = x_\pm$, $y
= y_\pm$.  Again they preserve a $U(1) \times U(1)$ symmetry generated
by (\ref{kAkB}) and the cosmological constant evaluated at those
points is given by ${\cal V} = - \frac14 (n -r_{\pm})^2 A_\pm^2$.  The
fact that must be pointed out is that now these critical points lie in
disconnected branches of our moduli space.  One can indeed see that
$x_+^2 + y_+^2 < 1$, whereas $x_-^2 + y_-^2 > 1$ and we can have at
most one critical point for each connected region.  This implies that
we cannot build any regular flow interpolating between them.


\subsection{A smooth domain wall solution}


As we mentioned before our non-homogeneous space has two physical
regions: $r \leq r_-$ and $r \geq {\rm max}\{n , r_+\}$.  
The situation in the latter case is very much similar to homogeneous
spaces, which is expected since this range survives the limit 
giving ${\mathbb E}AdS_4$ or ${SU(2,1) \over U(2)}$.
More interesting is the other physical region where $r \leq r_-$ and
let us discuss an explicit example.  

In the case that $\kappa = 1$, we have seen that we can have two
distinct critical points at $r = r_-$ (or equivalently $z = \bar z =
0$ in the coordinates for which the metric is well behaved), $x =
x_{\pm}$ and $y = y_\pm$.  As we have seen in (\ref{V+V-}), the value
of the potential, and therefore of the cosmological constant, at such
points differs in general.  This changes when the gauged symmetry is
completely inside the $SU(2)$ factor of the isometry group or it is
given by the extra $U(1)$.  Let us therefore choose in the following
the gauged isometry to be $k = 4 \sqrt{\frac23} \, T_1$, where the
coefficient has been fixed for later convenience.  This implies that
we will have two $AdS$ critical points with the same value of the
cosmological constant in the same region of the moduli space, and that
we can try to find the solution interpolating between them.

In the well behaved coordinates of (\ref{667}), the Killing vector 
corresponding to this isometry is given by
\begin{equation} 
\label{927}
 k = \sqrt{\frac83}\Big[ \, {\rm i} \, 4 n \, x \,  
 \beta_{\pm} (z \partial_z - \bar z 
 \partial_{\bar z}) + 2\, xy \,\partial_x + ( 1 - x^2 + y^2) \,
 \partial_y \Big] \ .
\end{equation} 
This gauging gives then two critical points at $x = \pm 1$, $y =0$ and
$z = \bar z = 0$.  At both critical points the cosmological constant
is ${\cal V} = -\frac14 (n - r_-)^2 (n+ r_+)^2$ and we are interested
in the solution of the flow equations (\ref{000}) interpolating
between them.  The corresponding superpotential $W$ can be obtained
from the Killing prepotentials (\ref{123}) using (\ref{Wmydef}).  For
any value of $x$ one can also see that $\partial_y W(z, \bar z, x, y)
= \partial_z W(z, \bar z, x,y) = \partial_{\bar z} W(z, \bar z, x, y)
=0$ if $z = \bar z = y =0$ and that therefore we can restrict to a
flow in the single $x$ coordinate, keeping fixed the others.  From
(\ref{000}) it follows indeed that in this case $y'=z'=\bar z' = 0$.
The superpotential restricted to this line is only a function of $x$
and reads
\be779
W =   \frac43 \; (n - r_-)(n+ r_+) \; {x \over 1 + x^2} 
\ee
with $r_{\pm} = - n \pm \sqrt{4n^2 -1}$.  
\begin{figure}[t]
    \centering
\includegraphics[angle=0,width=60mm]{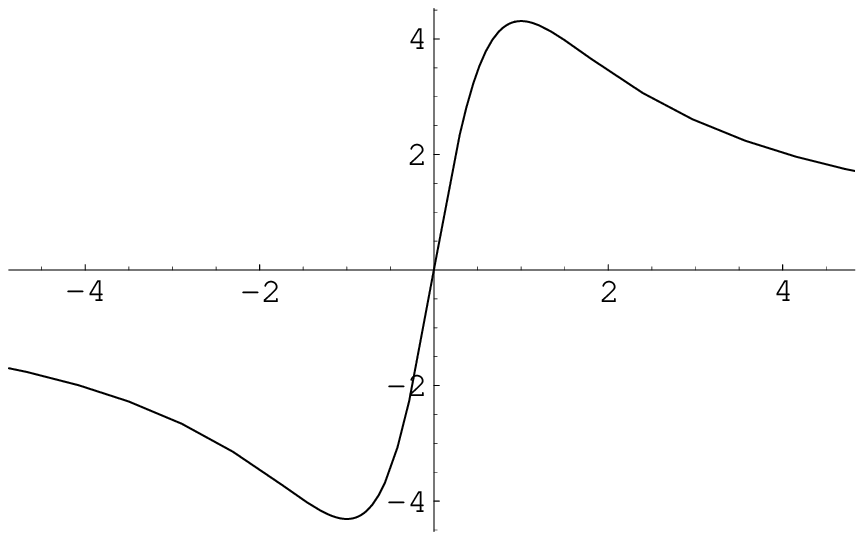}
\hspace{15mm} 
\includegraphics[angle=0,width=65mm]{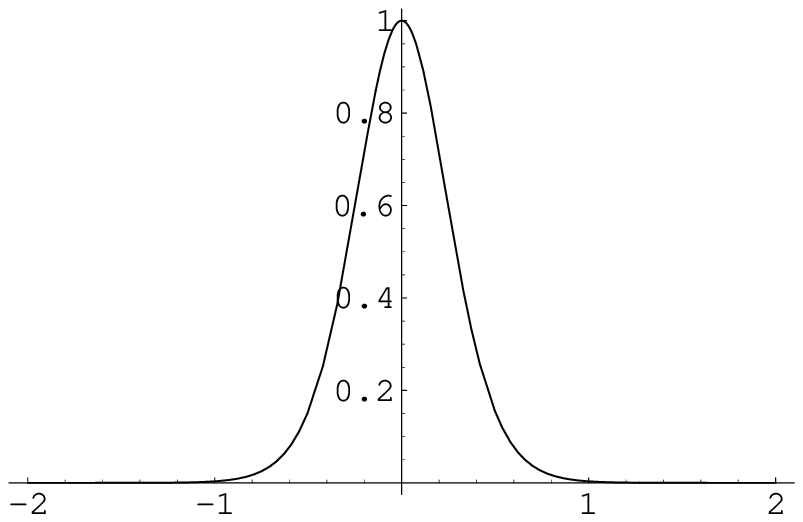} 
\caption{On
the left hand side we plotted the superpotential in the case $n=1$ as
a function of $x$ and the right hand side the solution for the warp
factor as function of $\rho$ interpolating between the two extrema} 
\label{Wfig}
\end{figure}
The flow equations for the scalar field $x$ and the warp factor $U$ become
\be723
\begin{array}{rcl}
x' &=& 3 \, g\, g^{xx} \partial_{x} W = \left(1- x^2 \right)\,, \\[3mm]
U' &=& -g\, W\,,
\end{array} 
\ee
whose solution is given by 
\be899
ds^2 = e^{2U} \, ds_{4d} + d\rho^2 \quad , \qquad 
x =  \tanh{(g \rho)}
\ee
with the warp factor 
\be772 e^{2U} = \left[\cosh\left(2 g \rho \right) \right]^{-\frac13
(n-r_-)(n+ r_+)} \ .  \ee The $AdS_5$ vacuum is reached at $\rho = \pm
\infty$ and since $(n-r_-)(n+ r_+)>0$ the warp factor vanishes
exponentially on both sides of the wall, see also the rhs in picture
(\ref{Wfig}).  This means that this domain wall solution can trap
gravity in the way suggested by Randall and Sundrum \cite{220}.
Moreover, as it can be seen from the picture (\ref{Wfig}), the
superpotential shows two extrema at $x =\pm 1$ and our solution
interpolates between them\footnote{One could ask what happens if the
scalar starts to roll down in the other direction, i.e. $x\to
\pm\infty$.  The answer would be that, despite the scalar is blowing
up, the solution would not change.  In fact the equations are
invariant under $x \rightarrow 1/x$ and the point $x = \pm \infty$ is
equivalent to the point $x =0$.  Notice, our coordinates do not cover
the point at $x \to \infty$, which are however good points of the
manifold (namely the north pole of the sphere).  Moreover, if one
thinks of our potential in the coordinates given in (\ref{angul}), it
becomes a periodic function ($W \sim \sin \theta$).}.  As expected for
brane--world type solutions, the superpotential $W$ crosses a point
where it vanishes, which is related through (\ref{000}) to the maximum
of the warp factor. Note, that in the so-called thin-wall limit of
large $g$ the scalar field becomes piecewise constant and the metric
approaches $AdS_5$ for any non-vanishing value of $\rho$.

Our result could seem in contrast with the no--go theorem in
\cite{210}, where the possibility of obtaining smooth Randall--Sundrum
type solutions in some generic classes of supergravity theories is
discussed.  But one of the assumptions for their result to hold is
that the potential should be non--positive ${\cal V} \leq 0$, at least
in the range of values of scalar fields that is explored in the
solution under consideration.  This is not the case in our solution.
Since the potential has a negative contribution coming from $W^2$ and
a positive one coming from its derivatives, there are points along the
solution where ${\cal V} >0$.  Take for instance the point $x = 0$,
where $W = 0$.  At that point $\partial_x W \neq 0$ (as the flows does
not stop there) and therefore the potential has a positive value.
This means that our example cannot be included into the class of
solutions considered in \cite{210} and therefore avoids the no--go
theorem presented there.


\section{Conclusion}


We discussed a class of 4--dimensional non--homogeneous quaternionic
spaces, which admit four isometries and have two disconnected physical
regions, where the metric is regular and positive definite.  Between
these two regions the space has timelike directions and exhibits a
curvature singularity.  This space is constructed as a non--trivial
line bundle over a 2--d base space of constant curvature $\kappa$. For
$\kappa =1$, this quaternionic space is the AdS-Taub-NUT solution with
a special value of the mass parameter and represents a resolution of
an orbifold of the Euclidean $AdS_4$ space. For vanishing NUT charge
the space becomes $AdS_4$ and in the limit of infinite NUT charge it
becomes ${SU(2,1) \over SU(2) \times U(1)}$. Therefore, it can be seen
as an interpolating solution between the two known homogeneous 4-d
quaternionic spaces.

As long as one considers the physical parameter range, where the
metric is positive definite, this space can be regarded in $N=2$ gauged
supergravity as the space parameterized by one hypermultiplet.  Doing
this, we discussed in the second part of the paper the isometries,
their gauging and the critical points of the resulting potential.  We
found for $\kappa =1$ that the critical points differ significantly in
the two allowed regions. In one region the situation is very much
similar to the gauging of the homogeneous spaces, i.e.\ there are no
multiple critical points (or critical surfaces) and hence there are no
smooth domain wall solutions. In the other region however, we found
two critical points with different values of the cosmological
constant, that are located exactly on the boundary of the physical
parameter range. The value of cosmological constant as well as the
nature of the critical points can continuously be changed while
varying the gauging parameters.  For a specific gauging we constructed
explicitly a smooth domain wall solution. Since this solution
interpolates between two IR critical points, the warp factor is
exponentially suppressed on both sides and thus gravity can be trapped
near this wall.

There are number of directions, which are worthwhile to pursue.  We
considered only an Abelian gauging of this space, but one may also
consider non-Abelian gaugings or discuss the gauging of two Abelian
isometries. In both cases one would have to couple this model also to
vector multiplets. This should be a straightforward generalization,
but more interesting would be to ask whether this space can appear in
compactified string- or M-theory and if yes, can one understand the
smooth domain wall in terms of branes.  Finally, it is needless to
say, that it would be interesting to construct also other
non-homogeneous quaternionic spaces and to discuss the resulting
vacuum structure (as e.g.\ the spaces discussed in \cite{333}).


\bigskip \bigskip

\noindent
{\bf Acknowledgments}

\medskip

We would like to thank Antoine van Proeyen for helpful
discussions. The work of K.B.\ is supported by a Heisenberg grant of
the DFG and G. D. acknowledges the financial support provided through
the European Community's Human Potential Programme under contract
HPRN-CT-2000-00131 quantum spacetime.




\providecommand{\href}[2]{#2}\begingroup\raggedright\endgroup



\end{document}